**Virtual boundary integral neural network for three-dimensional exterior acoustic problems**

Jiahao Li,[1] Qiang Xi,[1] Ilia Marchevskiy,[2,] and Zhuojia Fu[3,a]

[1] *College of Mechanics and Engineering Science, Hohai University, Nanjing 211100, China*

[2] *Department of Applied Mathematics, Bauman Moscow State Technical University, Moscow 105005, Russia*

This paper presents a virtual boundary integral neural network (VBINN) for exterior acoustic problems in three dimensions. The method introduces a virtual boundary inside the scatterer or vibrating body and represents the associated source density with a neural network. Coupled with the acoustic fundamental solution, this representation satisfies the Sommerfeld radiation condition by construction and enables direct evaluation of the acoustic pressure and its normal derivative at arbitrary field points. Because the integration surface is separated from the physical boundary, the formulation avoids the singular and near singular kernel evaluations associated with coincident source and collocation points in conventional boundary integral learning methods. To reduce sensitivity to boundary placement, the geometric parameters of the virtual boundary are optimized jointly with the source density during training. Numerical examples for acoustic scattering, multiple body interaction, and underwater acoustic propagation show close agreement with analytical solutions and COMSOL results, and the Burton Miller extension further improves stability near characteristic frequencies. These results demonstrate the potential of VBINN for exterior acoustic analysis in three dimensions.

[a] Email: paul212063@hhu.edu.cn

## I. INTRODUCTION

Three-dimensional acoustic radiation and scattering problems are important in many engineering applications, including noise control, acoustic design, and underwater acoustics[1]. Their numerical solution is challenging because they often involve complex geometries and unbounded domains, where the acoustic field must satisfy the Sommerfeld radiation condition at infinity. Conventional domain-based methods, such as the finite element method (FEM) and the finite difference method (FDM), are widely used for acoustic analysis and have shown strong performance for bounded-domain problems[2–4]. However, for exterior acoustic problems, these methods usually require truncation of the unbounded domain and the introduction of artificial boundary treatments, such as absorbing boundaries or perfectly matched layers. In three dimensions, these additional treatments complicate mesh generation and increase computational cost[5,6].

In recent years, physics-informed neural networks (PINNs) and related scientific machine learning methods have provided a new meshfree framework for solving partial differential equations[7–9]. By embedding the governing equations and boundary conditions in the loss function, these methods offer flexibility for both forward and inverse problems[10–12]. For exterior acoustic applications, however, standard PINNs still face important difficulties[13]. They typically require many collocation points in the computational domain[14,15]. This can make training expensive and unstable. The problem is more severe for three-dimensional cases. More importantly, standard PINNs do not naturally enforce the far field radiation condition. As a result, additional truncation boundaries or specially designed loss terms are often needed, which may reduce efficiency and introduce extra approximation errors.

For this reason, boundary integral equation (BIE) methods[16–18] are particularly attractive for acoustic radiation and scattering problems[19]. By using the fundamental solution, they transform domain problems into boundary integral formulations, reduce the problem dimensionality by one, and naturally satisfy the governing equation and the radiation condition in the exterior domain. These advantages have motivated recent efforts to combine BIEs with deep learning[20]. A variety of boundary integral based neural network models and boundary operator networks have been proposed[21]. These models have been applied to acoustic radiation and scattering and extended to more general PDE problems[22–24]. Examples include the boundary integral neural network[25,26], PIBI-Nets[27], and other related boundary based learning frameworks.

Despite this progress, singular integrals remain a major difficulty in neural network methods based on boundary integrals. When the source point approaches the physical boundary, the Green's function kernel becomes singular or nearly singular. In that case, accurate treatment usually requires specialized quadrature or regularization[28–30]. This increases both implementation complexity and training cost[31]. A possible way to avoid this issue is to place the source points on a virtual boundary separated from the physical boundary, as in the method of fundamental solutions and virtual boundary element methods[32–34]. In this way, the kernel is evaluated away from coincident source and boundary points. However, the performance of such methods is highly sensitive to the location of the virtual boundary. If the virtual boundary is too close to the physical boundary, the kernel may still be nearly singular. If it is too far away, the resulting system may become severely ill-conditioned[35]. In most existing approaches, the virtual boundary is chosen by empirical trial and error. A robust and adaptive selection mechanism is still lacking.

To address these issues, this paper proposes a Virtual Boundary Integral Neural Network (VBINN). Most existing boundary integral neural networks only learn unknown functions on the boundary. In contrast, VBINN learns the source density function on a virtual boundary. It can also learn the geometric parameters that define that virtual boundary. These parameters are optimized together during training. As a result, the method can automatically determine a virtual boundary that balances numerical stability and approximation accuracy.

The remainder of this paper is organized as follows. Section II presents the governing equations, the main methodology of VBINN, the Burton-Miller method, and the learning strategy for the virtual boundary. Section III presents a series of numerical examples, including acoustic scattering and underwater acoustic propagation problems, and compares the VBINN results with analytical solutions and COMSOL results. Finally, Section IV summarizes the main findings and discusses future work.

## II. METHODOLOGY

### A. Governing Equations for Acoustic Problems

The propagation of acoustic waves in an ideal fluid medium obeys the linear wave equation. In an ideal fluid, the complex amplitude of the time harmonic acoustic pressure $p(\mathbf{x})$ satisfies the Helmholtz equation in the unbounded exterior domain $\Omega$:

$$\Delta p(\mathbf{x}) + k^2 p(\mathbf{x}) = 0, \quad \mathbf{x} \in \Omega, \tag{1}$$

Here, $\mathbf{x}$ denotes the spatial position vector, $\Delta$ is the Laplace operator, and $k = \omega / c$ is the wavenumber, where $\omega$ is the angular frequency and $c$ is the sound speed in the fluid. To obtain a unique solution, boundary conditions are imposed on the scatterer boundary $\Gamma$. The typical boundary conditions considered in this paper include the Dirichlet and Neumann conditions:

$$p(\mathbf{x}) = g(\mathbf{x}), \quad \mathbf{x} \in \Gamma_D, \tag{2}$$

$$\frac{\partial p(\mathbf{x})}{\partial n} = i\rho\omega v_n(\mathbf{x}), \quad \mathbf{x} \in \Gamma_N, \tag{3}$$

where $\mathbf{n}$ is the unit outward normal vector pointing to the exterior computational domain, $\rho$ is the medium density, and $g(\mathbf{x})$ and $v_n(\mathbf{x})$ denote the prescribed boundary acoustic pressure and normal velocity distribution, respectively.

In addition, to characterize outward radiation in the unbounded exterior domain and exclude reflected waves at infinity, the acoustic field must satisfy the Sommerfeld radiation condition:

$$\lim_{r\to\infty} r\left(\frac{\partial p}{\partial r} - \mathrm{i}kp\right) = 0, \qquad r = |\mathbf{x}|. \tag{4}$$

## B. Issues in Deep Learning Methods for Exterior Acoustic Problems

To clarify the differences among PINN, BINN, and VBINN, we briefly summarize their loss constructions. Fig. 1 compares three neural network architectures.

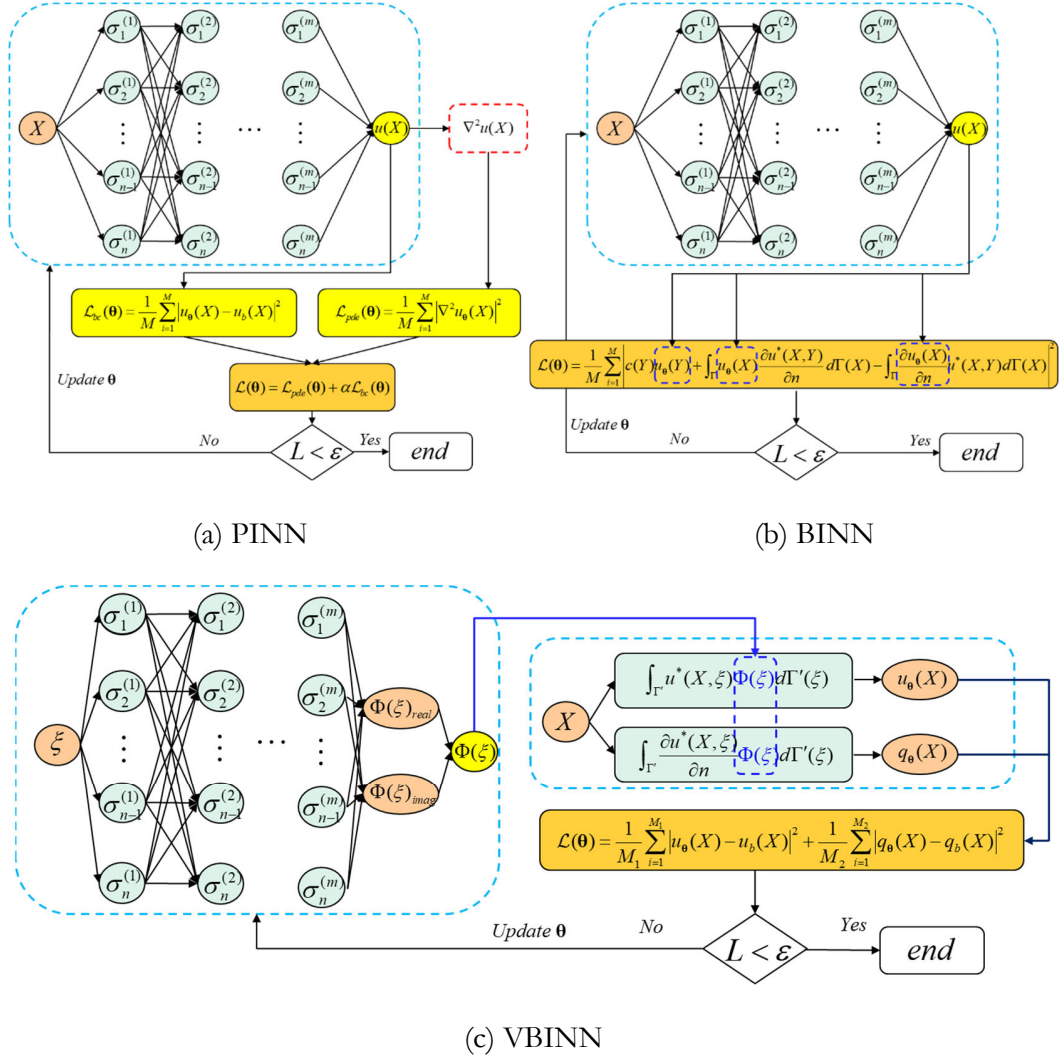

(a) PINN

(b) BINN

(c) VBINN

FIG. 1. Schematic comparison of the PINN, BINN, and VBINN architectures for the Helmholtz equation.

For exterior acoustic problems, a standard PINN introduces an artificial truncation boundary and augments the loss with an absorbing-boundary term, as illustrated in Fig. 1(a). Accordingly, the PINN loss function can be expressed as

$$L_{PINN} = L_f + \lambda_{bc} L_{bc} + \lambda_{abc} L_{abc}, \tag{5}$$

Here, $L_f$ denotes the PDE residual term, $L_{bc}$ denotes the term associated with the physical boundary conditions, and $L_{abc}$ denotes the absorbing boundary term imposed on the artificial truncation boundary. The parameters $\lambda_{bc}$ and $\lambda_{abc}$ are the corresponding weighting coefficients.

By contrast, boundary integral neural networks (BINN) reformulate the problem through boundary integral equations (BIEs), thereby avoiding artificial domain truncation and reducing the dimensionality of the problem, as shown in Fig. 1(b). For a point $\mathbf{x}$ on the physical boundary $\Gamma$, the standard Kirchhoff–Helmholtz integral equation can be written as:

$$c(\mathbf{x})p(\mathbf{x}) = \int_\Gamma \left( \frac{\partial p}{\partial n_\mathbf{y}} G(\mathbf{x},\mathbf{y}) - p(\mathbf{y}) \frac{\partial G(\mathbf{x},\mathbf{y})}{\partial n_\mathbf{y}} \right) dS_\mathbf{y}, \tag{6}$$

where $c(\mathbf{x})$ is a geometric coefficient.

Based on this, a BINN solves the acoustic field by minimizing the residual of the above boundary integral equation. Assuming the neural network outputs an approximate pressure $\hat{p}(\mathbf{x};\boldsymbol{\theta})$, the BINN loss is typically defined as the mean squared error over boundary collocation points $\{\mathbf{x}_i\}_{i=1}^{N_b}$:

$$L_{BINN} = \frac{1}{N_b} \sum_{i=1}^{N_b} \left\| c(\mathbf{x}_i)\hat{p}(\mathbf{x}_i) - \int_\Gamma \left( \frac{\partial \hat{p}}{\partial n_y} G(\mathbf{x}_i,\mathbf{y}) - \hat{p}(\mathbf{y}) \frac{\partial G(\mathbf{x}_i,\mathbf{y})}{\partial n_y} \right) dS_\mathbf{y} \right\|^2. \tag{7}$$

This architecture avoids domain truncation error. However, it still has important limitations. These limitations come from the mathematical nature of the integral term. During backpropagation for the gradient of Eq.(7), as the integration point **y** approaches the collocation point $\mathbf{x}_i$, the Green's function and its normal derivative exhibit strong singularities.

In summary, a new architecture is needed that satisfies the radiation condition as rigorously as integral equations while geometrically avoiding singularities. This is the fundamental motivation for moving the integration surface into the interior of the object and constructing a VBINN.

**C. VBINN Framework**

To address the above issues, this paper proposes the VBINN. The core idea is to move the integration surface into the interior of the object to construct an indirect representation of the solution, as illustrated in Fig. 1(c). The scattered acoustic pressure is then represented by

$$p^{sc}(\mathbf{x}) = \int_{\Gamma'} G(\mathbf{x},\mathbf{s})\phi(\mathbf{s})d\Gamma'(\mathbf{s}), \tag{8}$$

where $G(\mathbf{x},\mathbf{s})$ is the free space Green's function of the three-dimensional Helmholtz equation, and $\phi(\mathbf{s})$ is the unknown complex source density on $\Gamma'$. Since $\Gamma'$ is separated from the physical boundary $\Gamma$, the kernel is evaluated away from coincident points.

In VBINN, $\phi(\mathbf{s})$ is approximated by a neural network with parameters $\theta$. For each source point $\mathbf{s}_j \in \Gamma'$, the network outputs the real and imaginary parts of the density,

$$\phi_\theta(\mathbf{s}_j) = \text{Net}(\mathbf{s}_j;\theta)_1 + i\text{Net}(\mathbf{s}_j;\theta)_2. \tag{9}$$

The network is trained by enforcing the boundary conditions on the physical boundary $\Gamma$. For a Dirichlet problem, the pressure is prescribed on $\Gamma$, and the loss is defined as

$$\mathrm{L}_D(\theta) = \int_\Gamma \left| \int_{\Gamma'} G(\mathbf{x},\mathbf{s})\phi_\theta(\mathbf{s})d\Gamma'(\mathbf{s}) - g(\mathbf{x}) \right|^2 d\Gamma(\mathbf{x}). \tag{10}$$

For a Neumann problem, the normal derivative is prescribed on $\Gamma$, and the loss becomes

$$\mathrm{L}_N(\theta) = \int_\Gamma \left| \int_{\Gamma'} \frac{\partial G(\mathbf{x},\mathbf{s})}{\partial n_\mathbf{x}}\phi_\theta(\mathbf{s})d\Gamma'(\mathbf{s}) - v_n(\mathbf{x}) \right|^2 d\Gamma(\mathbf{x}), \tag{11}$$

where $g(\mathbf{x})$ and $v_n(\mathbf{x})$ denote the prescribed Dirichlet and Neumann boundary data, respectively.

In the numerical implementation, the continuous integrals are replaced by weighted summation over sampled points on $\Gamma$ and $\Gamma'$. The resulting discrete loss can be written as

$$L(\theta) = \frac{1}{N_b} \sum_{i=1}^{N_b} \left| \sum_{j=1}^{N_v} w_j K(\mathbf{x}_i, \mathbf{s}_j) \phi_\theta(\mathbf{s}_j) - b(\mathbf{x}_i) \right|^2, \quad (12)$$

where $\{\mathbf{x}_i\}_{i=1}^{N_b}$ are collocation points on the physical boundary, $\{\mathbf{s}_j\}_{j=1}^{N_v}$ are quadrature points on the virtual boundary, $w_j$ are the corresponding weights, and $K$ denotes $G$ for the Dirichlet case and $\partial G / \partial n_\mathbf{x}$ for the Neumann case. Accordingly, $b(\mathbf{x}_i)$ denotes $g(\mathbf{x}_i)$ or $v_n(\mathbf{x}_i)$.

With this construction, VBINN only requires boundary sampling. The acoustic field is expressed through a Green's function, which embeds the exterior governing equation and the relevant radiation condition into the solution form. Meanwhile, the separation between the virtual and physical boundaries keeps the kernel evaluation away from coincident points.

**D. Learning the Virtual Boundary Parameters**

The placement of the virtual boundary is crucial to the accuracy and numerical stability of the method. In conventional virtual boundary methods, its choice is constrained by two competing effects. If the virtual boundary $\Gamma'$ is placed too close to the physical boundary $\Gamma$, the kernel $G(x,s)$ becomes nearly singular as the source point $s$ approaches the field point $x$, thereby amplifying errors in numerical integration or discrete summation. In contrast, if $\Gamma'$ is placed too far from $\Gamma$, the basis functions associated with different source points tend to become linearly dependent, which results in a severely ill-conditioned algebraic system. For problems involving complex geometries or higher frequencies, this conflict is usually more pronounced, making the selection of virtual boundary parameters more sensitive.

To improve robustness, we introduce an $L_2$ regularization term on the source density and jointly optimize the source density network parameters and the virtual boundary radius. In implementation, the radius is parameterized as

$$R(\lambda) = R_{safe}\sigma(\lambda) = R_{safe}\frac{1}{1+e^{-\lambda}}, \tag{13}$$

where $\lambda \in \mathbb{R}$ is a trainable scalar and $R_{safe} = \eta R_{phy}$ is a prescribed upper bound determined from the scatterer geometry. This mapping ensures that the virtual boundary, parameterized here as a sphere, remains inside the scatterer, namely $R \in (0, R_{safe})$.

For the Neumann scattering case considered in this work, the training loss is defined as

$$\mathrm{L}(\theta,\lambda) = \frac{1}{N_b}\sum_{k=1}^{N_b}\left|\frac{\partial \hat{p}_{sc}(\mathbf{x}_k;\theta,\lambda)}{\partial \mathbf{n}} + \frac{\partial p_{inc}(\mathbf{x}_k)}{\partial \mathbf{n}}\right|^2 + \gamma\frac{1}{N_v}\sum_{j=1}^{N_v}\left|\phi_\theta(\mathbf{s}_j(\lambda))\right|^2, \tag{14}$$

where $\{x_k\}_{k=1}^{N_b}$ are collocation points sampled on the physical boundary $\Gamma$. The first term enforces the boundary condition, while the second penalizes excessively large source density and improves optimization stability.

**E. Burton-Miller Method**

In boundary integral modeling of acoustic scattering, classical integral equations may exhibit pronounced numerical instability at certain discrete frequencies and can even lead to non-unique exterior scattered field solutions; this phenomenon is commonly referred to as spurious characteristic frequencies. The primary cause is that when the operating frequency approaches the resonance frequencies of internal cavities, the coefficient system of the boundary integral equation becomes ill-conditioned, making the solution extremely sensitive to errors and resulting in instability or non-uniqueness.

To ensure uniqueness and stability across the full frequency band, this study introduces the Burton-Miller (BM) method into the VBINN framework. Instead of using a single boundary integral form, it employs a linear combination of the single-layer and double-layer potentials associated with the Green's function. The non-uniqueness of the single-layer form is typically associated with interior Dirichlet eigenvalues. The double-layer form is associated with interior Neumann eigenvalues. The problematic frequencies for these two forms generally do not coincide. By combining the two representations, one can effectively avoid singular behavior of a single form at specific frequencies and theoretically guarantee uniqueness of the solution for the exterior acoustic field.

In the VBINN representation used in this work, the scattered pressure is obtained by superposition of contributions from multiple virtual source points $\mathbf{s}_j$. For an arbitrary field point $\mathbf{x}$, the predicted scattered pressure is written in the following Burton-Miller form:

$$\hat{p}_{sc}(\mathbf{x};\theta) = \sum_{j=1}^{N_v} w_j \left[ G(\mathbf{x},\mathbf{s}_j) + i\alpha_{BM} \frac{\partial G(\mathbf{x},\mathbf{s}_j)}{\partial n_{s_j}} \right] \phi_\theta(s_j(\lambda)), \tag{15}$$

where $G(\mathbf{x},\mathbf{s}_j)$ is the free field Green's function, $w_j$ denotes numerical weights, $\phi(\mathbf{s})$ is the source density function output by the network at the virtual source points, $\mathbf{n}_{s_j}$ is the unit normal at the virtual source points, and $\alpha_{BM}$ is the coupling coefficient.

Correspondingly, the normal derivative commonly used in boundary conditions can be written as:

$$\frac{\partial \hat{p}_{sc}(\mathbf{x};\theta)}{\partial n_{\mathbf{x}}} = \sum_{j=1}^{N_v} w_j \left[ \frac{\partial G(\mathbf{x},\mathbf{s}_j)}{\partial n_{\mathbf{x}}} + i\alpha_{BM} \frac{\partial^2 G(\mathbf{x},\mathbf{s}_j)}{\partial n_{\mathbf{x}} \partial n_{s_j}} \right] \phi_\theta(s_j(\lambda)). \tag{16}$$

Moreover, the BM formulation involves the second-order normal derivative of the kernel, which is hypersingular in conventional boundary integral formulations when the source and collocation points coincide on the boundary. In VBINN, however, the virtual source points are placed inside the object boundary, while the collocation points remain on the physical boundary. Therefore, the first-and second-order derivative kernels remain nonsingular in the present formulation, and no special treatment of singular or hypersingular integrals is required. In the present work, the BM coupling parameter is taken as $\alpha_{BM} = 1/k$.

## III. NUMERICAL EXAMPLES

### A. Plane-wave scattering by a unit sphere

This example aims to verify the numerical accuracy and stability of the VBINN in solving three-dimensional acoustic scattering problems with Neumann boundary conditions. Consider a rigid sphere of radius $a = 1.0$ m placed in an infinite fluid medium with sound speed $c = 343$ m/s and fluid density $\rho_f = 1.225$ kg/m$^3$. A plane acoustic wave is incident along the positive $x$-axis. The corresponding scattering configuration is illustrated in Fig. 2. The incident pressure field is $p^{inc} = e^{ikx}$.

Since the sphere surface satisfies the Neumann boundary condition, i.e., the normal particle velocity on the boundary is zero, the total acoustic pressure satisfies the following boundary condition:

$$\frac{\partial p_{total}}{\partial n} = \frac{\partial (p_{inc} + p_{sc})}{\partial n} = 0, \quad \text{on } \Gamma, \tag{17}$$

where $n$ is the unit outward normal vector on the boundary. The analytical solution of this problem is given below and is used to validate the accuracy of the numerical solution:

$$p_{total}(r,\theta) = \sum_{n=0}^{\infty} i^n (2n+1) \left( j_n(kr) - \frac{j_n'(ka)}{h_n^{(1)'}(ka)} h_n^{(1)}(kr) \right) P_n(\cos\theta), \tag{18}$$

Here, $j_n$ and $h_n^{(1)}$ are the spherical Bessel function and the spherical Hankel function of the first kind, respectively, and $P_n$ is the Legendre polynomial.

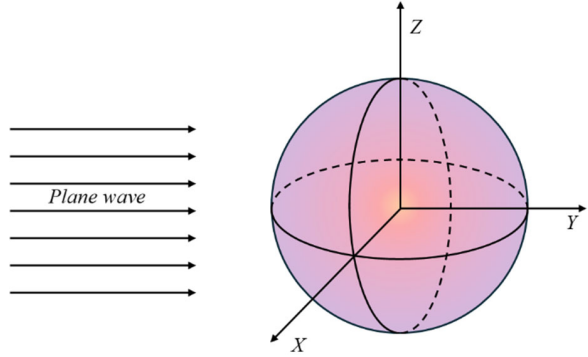

FIG. 2. Schematic of plane wave incidence on a rigid sphere

Fig. 3 presents contour maps of the total pressure magnitude around the rigid sphere for an incident wavenumber of $1.0 \text{ rad/m}$. The VBINN result in Fig. 3(a) matches the analytical solution in Fig. 3(b) very well. Fig. 3(c)–3(e) show that VBINN gives a smaller relative error than BINN and PINN. Fig. 3(f) shows that VBINN converges faster and reaches a lower loss value during training.

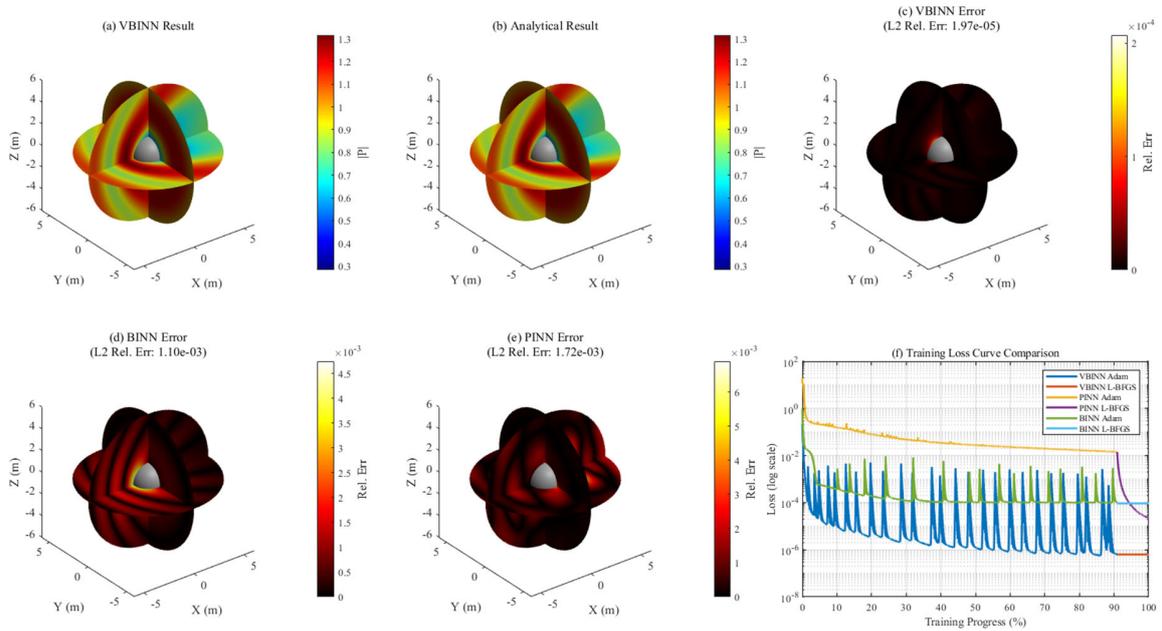

FIG. 3. Comparison of field distributions and relative error distribution

To complement the accuracy comparison, Table I reports the training cost of PINN, BINN, and VBINN for the same benchmark. PINN requires many interior collocation points and more optimization steps, which leads to a much higher training cost. BINN and VBINN are much more efficient. Although VBINN takes slightly longer to train than BINN in this test, it achieves higher accuracy.

TABLE I. Training cost comparison of PINN, BINN, and VBINN for the spherical scattering benchmark.

| Method | N | Adam | $\eta_{Adam}$ | $L-BFGS$ | $\eta_{L-BFGS}$ | $Time(s)$ |
|--------|------|------|-----------|---------|------------|---------|
| PINN   | 17400 | 5000 | $10^{-3}$ | 1000    | 1.0        | 1466.90 |
| BINN   | 1284 | 5000 | $10^{-3}$ | 500     | 1.0        | 16.08   |
| VBINN  | 1200 | 5000 | $10^{-3}$ | 500     | 1.0        | 22.31   |

To evaluate the convergence performance of VBINN, we fix the incident wavenumber at $k = 1.0$ rad/m and investigate the effect of the number of boundary sampling points $N$ on computational accuracy.

The numerical simulation results are shown in Fig. 4. As the number of boundary sampling points increases from 100 to 800, the model's relative $L_2$ error decreases significantly. This result indicates that VBINN converges steadily as the number of boundary sampling points increases, while maintaining a relatively low computational cost even under high accuracy requirements.

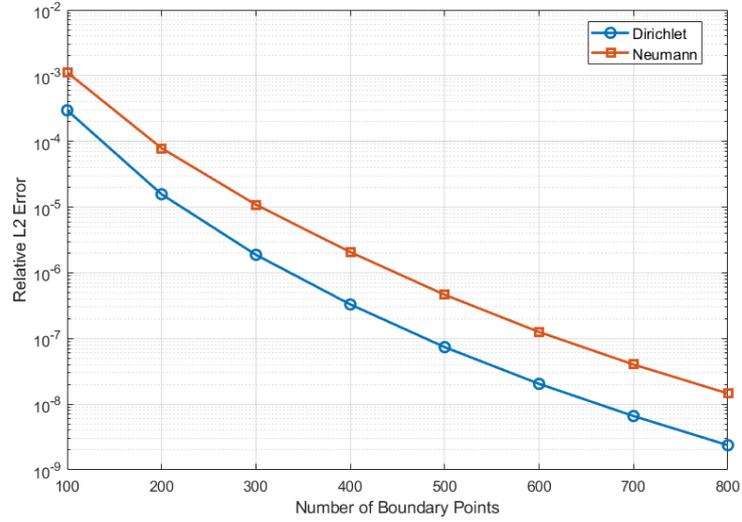

FIG. 4. Convergence of the VBINN under Dirichlet and Neumann Boundary Conditions

Next, we introduce the BM method into VBINN and examine its effect on frequency stability. Figure 5 shows the real and imaginary parts of the scattered pressure at an observation point on the sphere surface for $k \in [1,8]$ rad/m. With the BM method, the VBINN results remain smooth over the whole wavenumber range and agree well with the analytical solution. Without this treatment, visible oscillations appear near some characteristic frequencies. This result shows that the BM method improves stability in the present study.

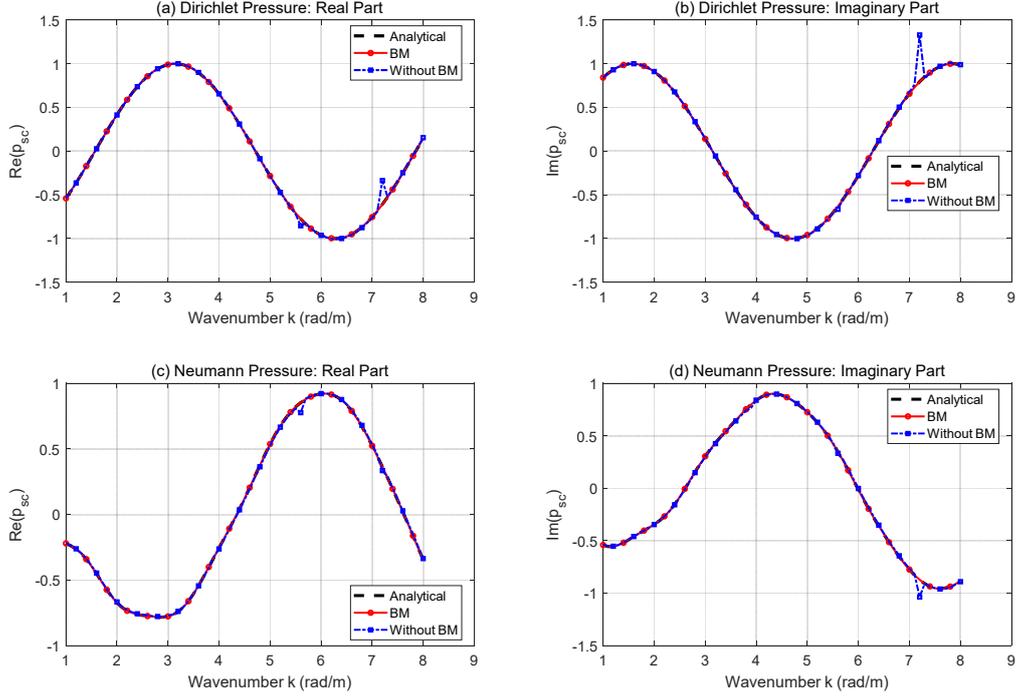

FIG. 5. Comparison of pressure responses before and after introducing the Burton-Miller method.

**B. Pea shaped scatterer**

This example further investigates the applicability of the method to irregular geometries and illustrates the adaptive virtual boundary strategy through the evolution of the learned radius.

We consider a three-dimensional pea shaped scatterer with a smooth but irregular surface, as shown in Fig. 6. Its geometry is defined by

$$F(x,y,z) = \frac{x^2}{0.64\left(1-0.1\cos\left(\frac{\pi z}{R}\right)\right)} + \frac{\left(y+0.3\cos\left(\frac{\pi z}{R}\right)\right)^2}{0.64\left(1-0.4\cos\left(\frac{\pi z}{R}\right)\right)} + z^2 = R^2. \quad (19)$$

where $R$ is the geometric parameter that controls the overall size of the scatterer.

This example adopts the method of manufactured solutions to validate the numerical accuracy of the algorithm. Assume that the acoustic field is generated by a point source located at the origin, so that the analytical solution is $p(r) = e^{ikr}/r$. The medium wavenumber is set to $k = 1.0 \text{ rad/m}$. A Neumann boundary condition consistent with this analytical result is imposed on the boundary $\Gamma$; VBINN is then used to reconstruct the exterior acoustic field, and the reconstruction is compared with the analytical result.

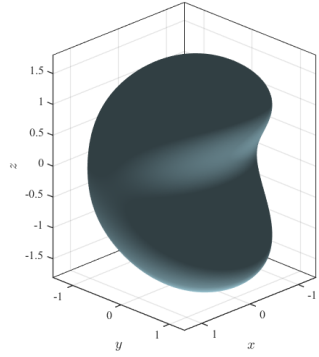

FIG. 6. Geometry of the pea shaped scatterer.

Figure 7 shows that VBINN accurately recovers the magnitude as well as the real and imaginary parts of the field. The agreement remains good even in the concave region, where the wave pattern is more intricate.

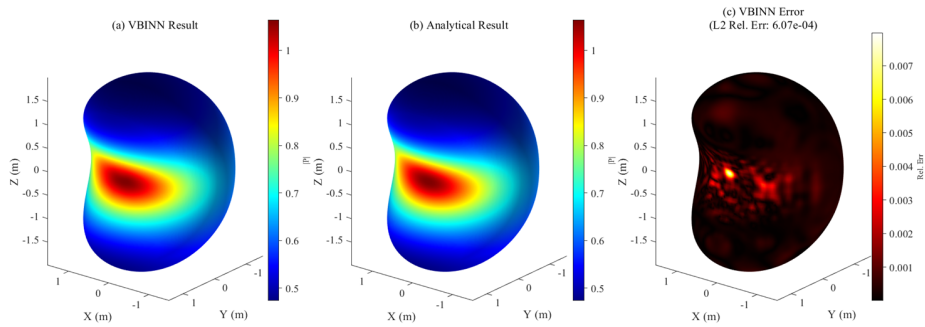

FIG. 7. Reconstructed acoustic field and relative error distribution for the pea shaped scatterer.

Introducing a learnable virtual boundary radius parameter provides a practical way to reduce manual parameter tuning in conventional virtual boundary element methods. In this study, the loss function contains a weighted combination of a data fitting term and a source density regularization term. This allows the virtual boundary location to evolve automatically during training.

Fig. 8(a) shows how the virtual sphere radius $R$ evolves during training under different initial radius coefficients. The black dashed line marks the minimum distance from the true physical boundary to the origin, with $R_{min} = 1.000$. It is clear that, whether the initial radius is set to $0.2$ or $0.94$, $R$ eventually converges to a stable value close to $0.89$. This suggests that the adaptive mechanism is insensitive to the initial radius and consistently converges to a stable value.

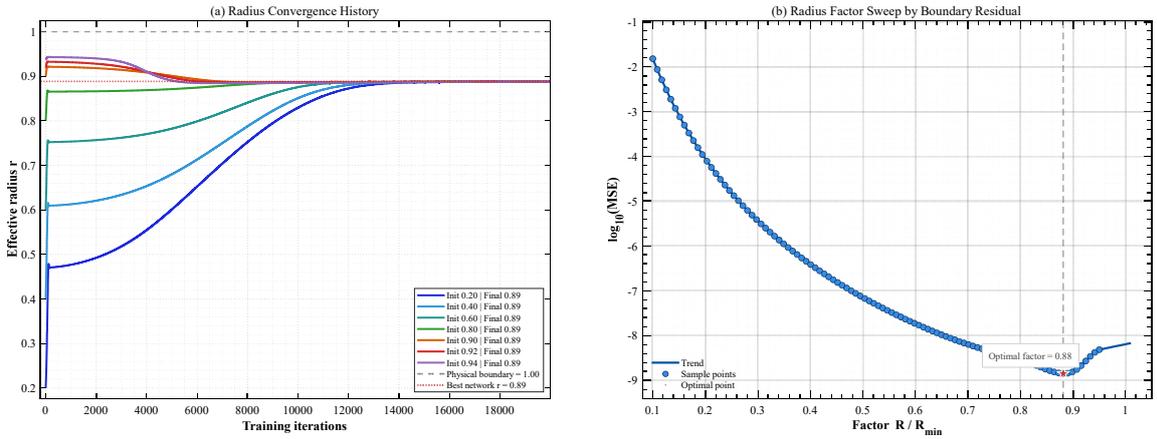

FIG. 8. Convergence of the virtual boundary radius under different initial conditions.

To further show that this converged value is reasonable, we also performed a global sweep over the virtual sphere radius using the conventional virtual boundary element method. As shown in Fig. 8(b), the optimal radius is found to be $R = 0.88$ by comparing the mean squared errors at different radius values. This result is very close to the radius identified by VBINN, indicating that VBINN can identify a suitable virtual boundary location without an exhaustive parameter sweep.

### C. Acoustic scattering by four spheres

In this example, we consider acoustic scattering by a system of four identical spheres and examine its far-field directivity under different incident conditions. Each sphere is treated as an ideal scatterer with radius $R = 0.4$ m, and the surrounding medium is air with sound speed $c = 343$ m/s. A Dirichlet boundary condition is imposed on each sphere surface so that the total acoustic pressure vanishes on the boundary. Within the VBINN framework, one virtual sphere is placed inside each physical sphere, with the virtual radius set to $R_{\text{virtual}} = 0.3$ m. The geometric arrangement of the four-sphere system is shown in Fig. 9.

Two incident conditions are considered. The first is normal incidence, where the plane wave propagates along the positive $z$-direction. The second is oblique incidence at an angle of $\pi/3$, where the incident wave lies in the $xz$-plane and is tilted by $\pi/3$ from the positive $z$-direction. For each incident condition, two dimensionless wavenumbers, $kR = 1$ and $kR = 4$, are considered to examine the scattering behavior of the array at relatively low and higher frequencies.

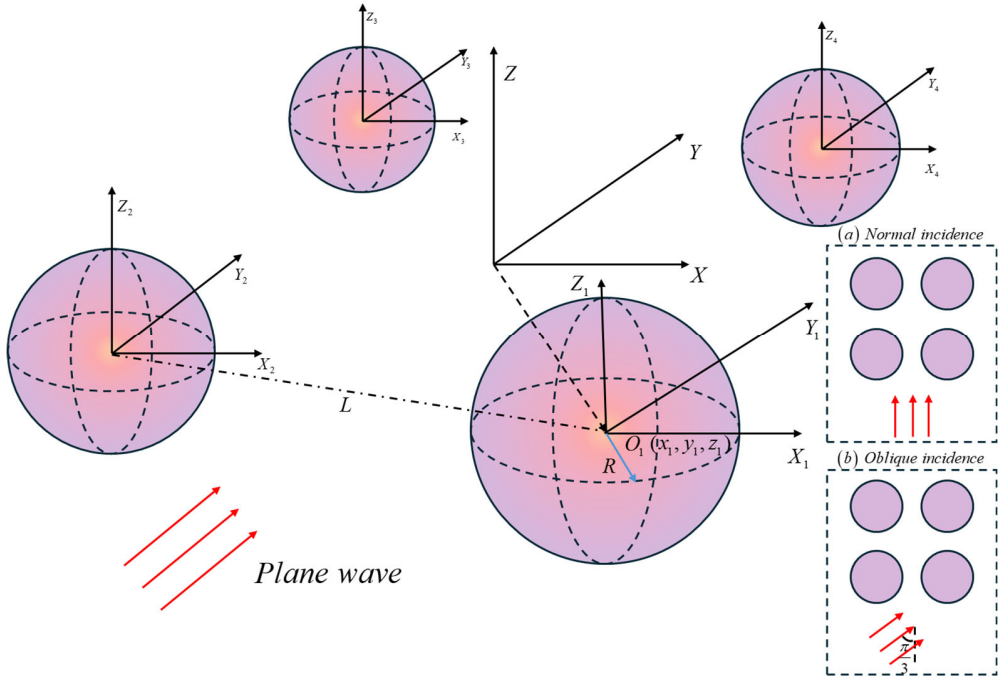

FIG. 9. Geometry of the four spheres scattering system.

Figure 10 illustrates the far-field scattering directivity of the four-sphere array for different incident conditions. As $kR$ increases, the main lobe becomes narrower and its peak becomes higher, while additional side lobes emerge and the angular pattern becomes more oscillatory. This indicates that the scattering field exhibits stronger directional selectivity and more pronounced multi-sphere interference at higher frequency. In addition, increasing $L/R$ generally strengthens the dominant scattering peak and makes the directional pattern more structured. For oblique incidence at $\pi/3$, the main lobe shifts relative to the normal-incidence case, although the overall trends with respect to $kR$ and $L/R$ remain unchanged. These results show that VBINN can accurately capture the dependence of multi-body far-field scattering on frequency, spacing, and incident direction.

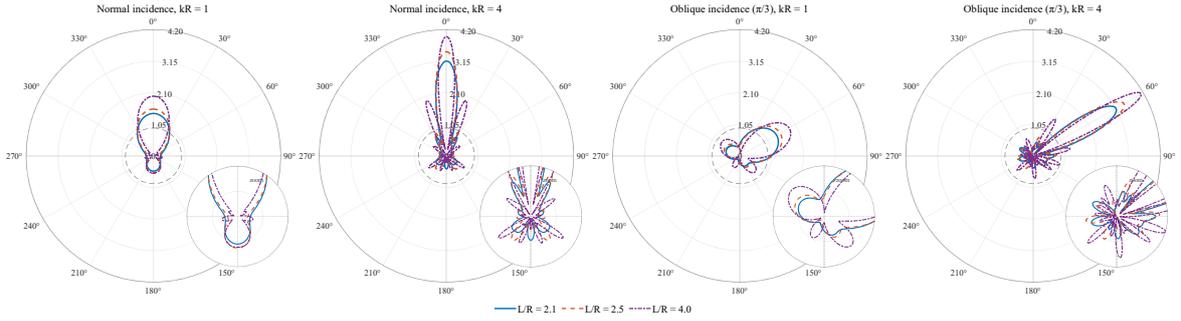

FIG. 10. Far-field directivity of the four-sphere array for normal incidence and oblique incidence at $\pi/3$: (a) normal incidence, $kR = 1$; (b) normal incidence, $kR = 4$; (c) oblique incidence at $\pi/3$, $kR = 1$; and (d) oblique incidence at $\pi/3$, $kR = 4$.

### D. Underwater acoustic propagation from a vibrating capsule shell

In the final example, we study underwater sound propagation generated by a vibrating capsule shell in a shallow ocean environment. The capsule shell has a thickness of $th = 1$ mm, and a horizontally oriented harmonic outward distributed force with amplitude $F_3 = 5$ N/m is applied on its outer surface; the geometry and loading are shown in Fig. 11. The shallow ocean water depth is $H = 15$ m, and the shell is submerged to a depth of $h = 7$ m. In this example, the numbers of virtual boundary points and training points are set to $N_v = N_b = 100$.

Different from the free-space exterior problems in Examples A–C, the shallow-ocean case is modeled with an environment-dependent Green's function. In this case, the kernel $\mathbf{K}$ in Eq. (12) is replaced by the corresponding ocean Green's function that satisfies the surface and bottom boundary conditions, while the VBINN training procedure remains unchanged.

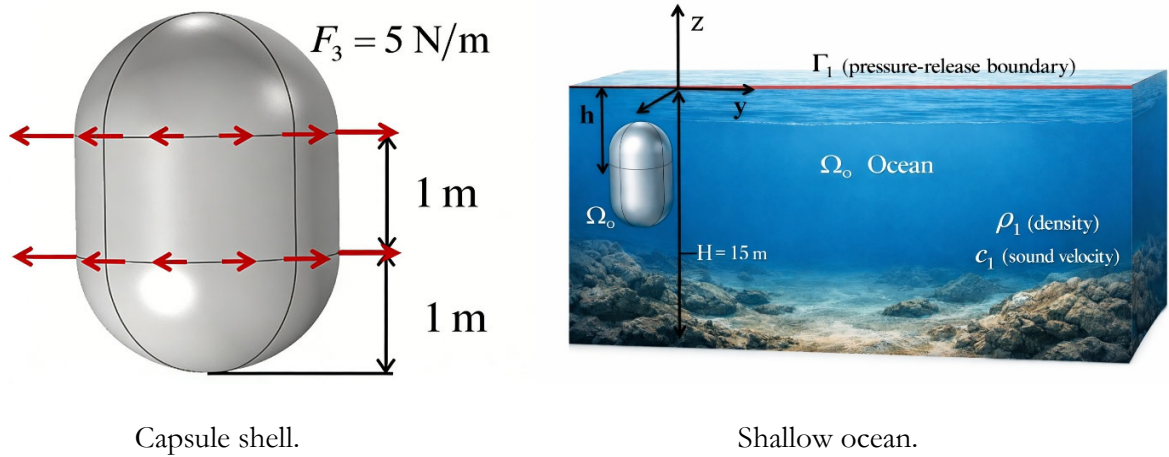

Capsule shell.  Shallow ocean.

FIG. 11. Geometric model of the capsule shell and shallow ocean.

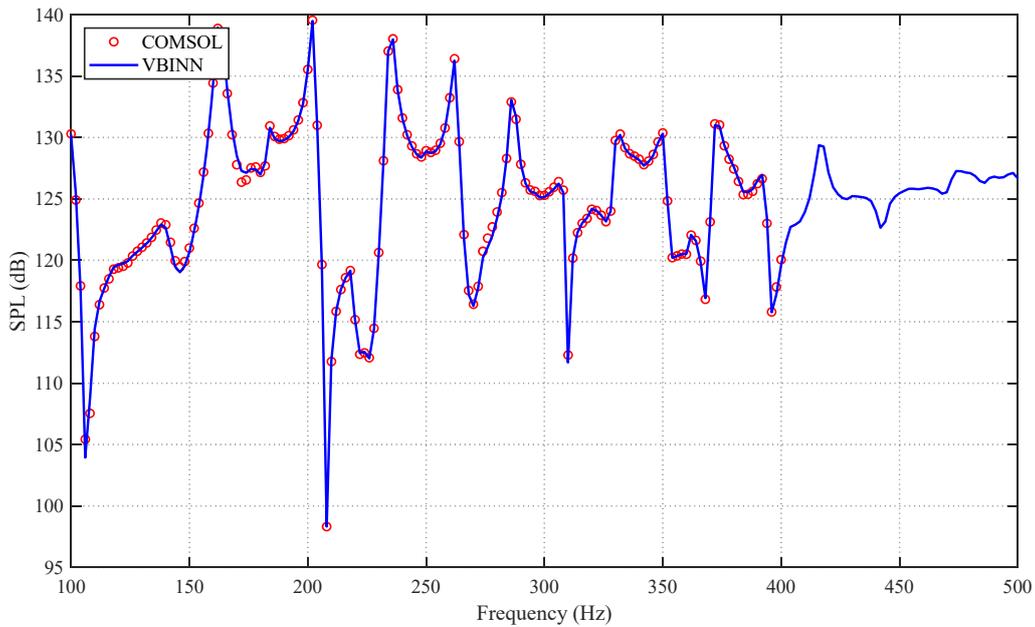

FIG. 12. Comparison of SPL in the shallow ocean environment predicted by VBINN and COMSOL.

Fig. 12 presents the frequency response curves of the underwater sound pressure level (SPL) at the observation point $(x, y, z) = (4\text{m}, 0\text{m}, -7\text{m})$. The SPL curves predicted by VBINN agree well with the COMSOL results in the frequency range up to 400 Hz. It should also be noted that, on an ordinary personal computer, the COMSOL simulation in the present example can be carried out up to about 400 Hz, but the computational time becomes increasingly long as the frequency increases. By comparison, VBINN can further extend the prediction to 500 Hz. This indicates that the proposed method remains applicable for higher-frequency response analysis in the present shallow-ocean problem.

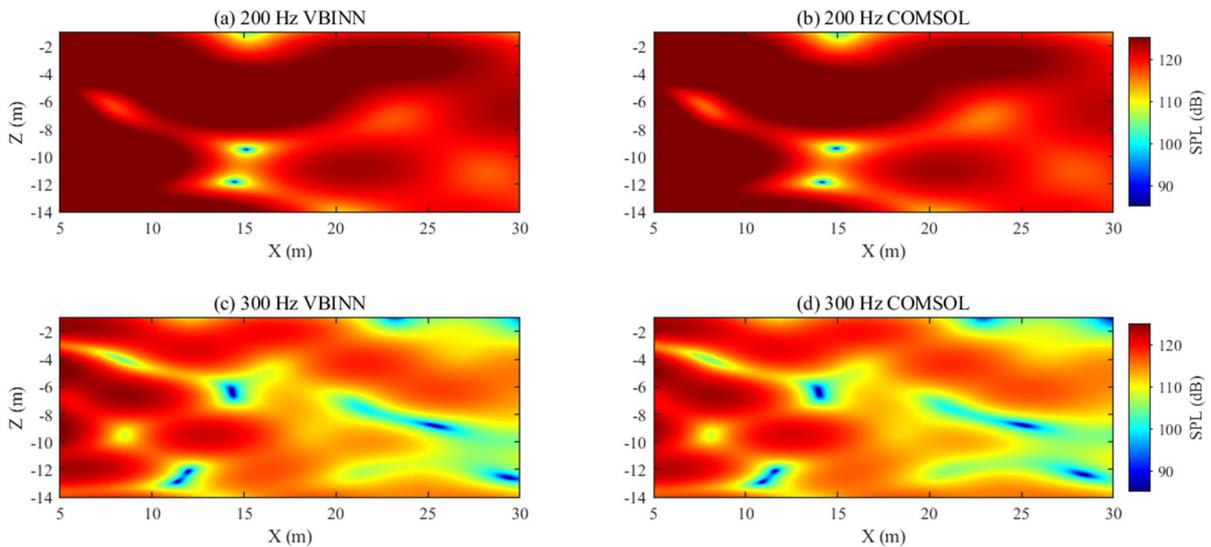

FIG. 13. Spatial distribution of SPL for the capsule shell case.

Furthermore, Fig. 13 shows the spatial distribution of the sound pressure level at an excitation frequency of 300 Hz. The spatial distributions predicted by VBINN also agree well with the COMSOL results. These comparisons validate the effectiveness of VBINN for predicting underwater acoustic propagation in shallow-ocean environments.

## IV. CONCLUSION

This paper proposed a virtual boundary integral neural network (VBINN) for three-dimensional exterior acoustic problems. By introducing a virtual boundary inside the scatterer or vibrating body, the proposed method avoids the singular kernel evaluations associated with coincident source and collocation points in conventional boundary integral learning methods. In addition, the virtual boundary parameters are optimized together with the source density representation during training, which reduces the dependence on empirically chosen boundary locations.

The numerical examples demonstrated the accuracy, stability, and applicability of the proposed method. For plane wave scattering by a unit sphere, VBINN showed close agreement with the analytical solution and achieved higher accuracy than BINN and PINN while maintaining much lower computational cost than PINN. For the irregular pea-shaped scatterer, VBINN accurately reconstructed the exterior acoustic field, and the learned virtual boundary radius converged to a value close to the optimum obtained by a global sweep. For the four-sphere system, the method successfully captured the dependence of far-field directivity on frequency, spacing, and incident direction. For the shallow ocean capsule shell problem, the predicted SPL responses and spatial distributions agreed well with the COMSOL results. The BM method further improved the stability near characteristic frequencies. Overall, the present results show that VBINN is an accurate and stable method for three-dimensional exterior acoustic analysis.

Future work may focus on improving the computational efficiency for large-scale and multiscale problems by incorporating mature acceleration techniques such as the fast multipole method. In addition, since the present model is trained for a single geometry or a single frequency setting, future research may explore operator learning frameworks incorporating Green's function priors to improve generalization across geometries and frequency conditions.

**ACKNOWLEDGMENTS**

The work reported in this paper was supported by the Natural Science Foundation of China (Grant Nos. 12372196, 12302258), the Fundamental Research Funds for the Central Universities (Grant No. B250201051).

# REFERENCES (NUMERICAL STYLE)